\documentclass[a4paper,fleqn,usenatbib]{mnras}

\usepackage{newtxtext,newtxmath}
\usepackage[T1]{fontenc}
\usepackage{ae,aecompl}

\usepackage{graphicx}	% Including figure files
\usepackage{amsmath}	% Advanced maths commands
\usepackage{amssymb}	% Extra maths symbols
\usepackage{multicol}
\usepackage{ulem,color}

\usepackage{caption} 
\title[Mass ejection in Short GRBs]{Observational evidence  for  mass ejection accompanying short gamma ray bursts }

\author[Moharana \& Piran]{
Reetanjali Moharana,$^{1}$\thanks{E-mail:moharana.reetanjali@mail.huji.ac.il}
and Tsvi Piran$^{1}$
\\
$^{1}$Racah Institute of Physics, The Hebrew University, Jerusalem 91904, Israel
}

\pubyear{2017}

\begin{document}
\label{firstpage}
\pagerange{\pageref{firstpage}--\pageref{lastpage}}
\maketitle

% Abstract of the paper
\begin{abstract}
The plateau in the duration distribution of long Gamma-Ray Bursts (LGRBs) provides a  direct observational evidence for the Collapsar model. The plateau reflects the fact that the observed duration satisfies: $T_{90} = 
t_{e}-t_{b}$ where $t_{e}$ is the time that the central engine operates and $t_{b}$ is a threshold time,  interpreted within the Collapsar model as the time it takes for the relativistic jet  to penetrate the stellar envelope. Numerical simulation and macronova observations suggest that compact binary mergers involve mass ejection.
If 
short-Gamma Ray Bursts (sGRBs) arise from such mergers,  their jets should  cross this surrounding ejecta
before producing the prompt emission.  Like in  LGRBs,  this  should result in a distinct short plateau in the GRBs'  duration distribution. We  present a new analysis of  the duration distribution for the three GRB satellites: BATSE, {\it Swift} and Fermi. We find  a clear evidence for a short  ($\sim 0.4$ sec) plateau in the duration distribution. This  plateau is  consistent with the expected jet crossing time, provided that the ejecta is of order of a few percent of solar masses. 
\end{abstract}

% Select between one and six entries from the list of approved keywords.
% Don't make up new ones.

\maketitle
\flushbottom
\section{Introduction}
\label{intro}
Gamma ray bursts (GRBs) are broadly divided into two distinct groups: long  (LGRBs) with $T_{90}$> 2 sec{\footnote{$T_{90}$ is the duration in which the central 90\% of the gamma-ray signal is detected.}} and short  (sGRBs) with $T_{90}$ < 2 sec. The association of long GRBs with star forming regions \citep{pac98} that was followed by the discovery, in 1998,   \citep{galama98,Iwamoto:1998tg} %,E.:2015tia}
of Supernovae (SNe)-GRB associations revealed that  LGRBs are related to the death of massive stars. 
The origin of sGRBs remained obscure until 2005, when the  {\it Swift}  and HETE-2 satellites detected the first sGRB afterglows from GRBs 050509b, 050709 and 050724~ \citep{Fox2005,Gehrels:2005qk,Tirado2005} leading to the identification of their host galaxies. The observational difference between the host Galaxies,  LGRBs host have  large star-formation rate while  sGRB hosts consist of both star forming and non-starforming galaxies  \citep{Barthelmy:2005bx}, suggests that  sGRBs result from different  progenitors than LGRBs  \citep[see e.g.][]{Nakar:2007yr}. Recent observations of 
macronovae/kilonovae  \citep{Li:1998bw,Metzger2010,Kasen:2013xka,Barnes:2013wka,Hotokezaka2013} associated with some sGRBs: 130603B  \citep{Berger2013,tanvir2013}; 060614  \citep{Yang:2015pha} and 050709  \citep{Jin:2015dxh} support earlier theoretical predictions 
 \citep{Eichler:1989ve}
 %,Narayan:1992iy} 
 that  sGRBs arise during compact binary mergers, either a neutron star-neutron star (NS-NS), or a neutron star - Black hole (NS-BH) mergers. 

The theoretical framework of LGRB-SNe association, the Collapsar model,  \citep{MacFadyenWoosley98} suggests that a
central engine generates a relativistic jet within the core of the collapsing star.  The jet propagates with a  head velocity of $\sim0.1c$.  The  GRB is  produced after the jet emerged from the surface of the collapsing star. \cite[][hearafter B12]{bromberg12} provided a direct observational confirmation of  this model. 
While the jet is within the star  the central engine must be active in order that the jet continues to propagate. 
Therefore, the duration of the prompt emission, $T_{90}$, is  the time that the engine operates, $t_{e}$ after the jet breakout time, $t_{b}$:~ $T_{90}=t_{e}-t_{b}$. This last relation gives rise to a plateau feature in the LGRB  duration distribution ($dN_{GRB}/dT_{90} \approx {\rm const.}$) for $T_{90}\leqslant t_{b}$  \cite[][hearafter B13]{bromberg13} . This plateau is indeed observed in the LGRB $T_{90}$ distribution and its length, 
$t_{b} \sim 15$ secs, is consistent with theoretical expectations  \citep{bromberg11}.

Compact binary mergers are  accompanied by substantial amount of dynamical mass ejection  \citep[see e.g.][and references therein]{Hotokezaka2015}. The excess in near-IR band observed by { Hubble space Telescope} in {\it Swift}  SGRB 130603B  \citep{Berger2013,tanvir2013} is explained by the macronova model provided that a large amount of mass $\gtrsim 2 \times 10^{-2} M_{{\odot}}$ is ejected in the NS-NS merger and is powered by the radioactivity of r-process nuclei  \citep{Hotokezaka:2013kza,Berger2013,tanvir2013}. Even larger amount of mass ejection have been suggested in other cases ( $\sim0.05 M_{\odot}$ for GRB 050709  \citealt{Jin:2015dxh};  and $\sim 0.13 M_{\odot}$ for GRB 060614  \citealt{Yang:2015pha}). Hence like in the Collapsar model, the relativistic jet in a merger  propagates through an expanding merger ejecta of a significant mass  \citep{Nagakura:2014hza,Murguia-Berthier:2014pta}. Therefore,  if the progenitors of sGRBs are compact binary mergers  we should expect a similar plateau, like the one in LGRBs but shorter, in the shorter part of the  duration distribution  reflecting the typical time it takes for a merger's jet to reach  the ejecta edge.

In this letter, we analyzed the data for the GRB duration distribution following  B12,B13  for the three major GRB satellites, BATSE, {\it Swift}  and Fermi-GBM. 
However, unlike these earlier papers that focused  on the LGRB time scale we explore here also the possibility of a plateau in the short duration part of  the $T_{90}$ distribution.  We compare,  using the unbinned-maximum likelihood method,  the whole GRB duration distribution for each one of these detectors with a combined model featuring a ``long" plateau  for  Collapsars and a ``short" plateau for mergers.  
We organized the paper as follow, section ~\ref{theory} describe briefly an updated estimate of the jet penetration time in the case of a merger ejecta. In section~\ref{Data} we discuss the data sets from the three GRB detectors:
BATSE, {\it Swift} and Fermi-GBM.  We describe the the likelihood analysis method in section~\ref{likeli_method} and the results in section~\ref{ana_result}.  We conclude in section~\ref{discussion_con} with a summary of  the results and their  implications.

\begin{figure*}
 \begin{multicols}{2}
  \centering
 \includegraphics[width=1.\linewidth]{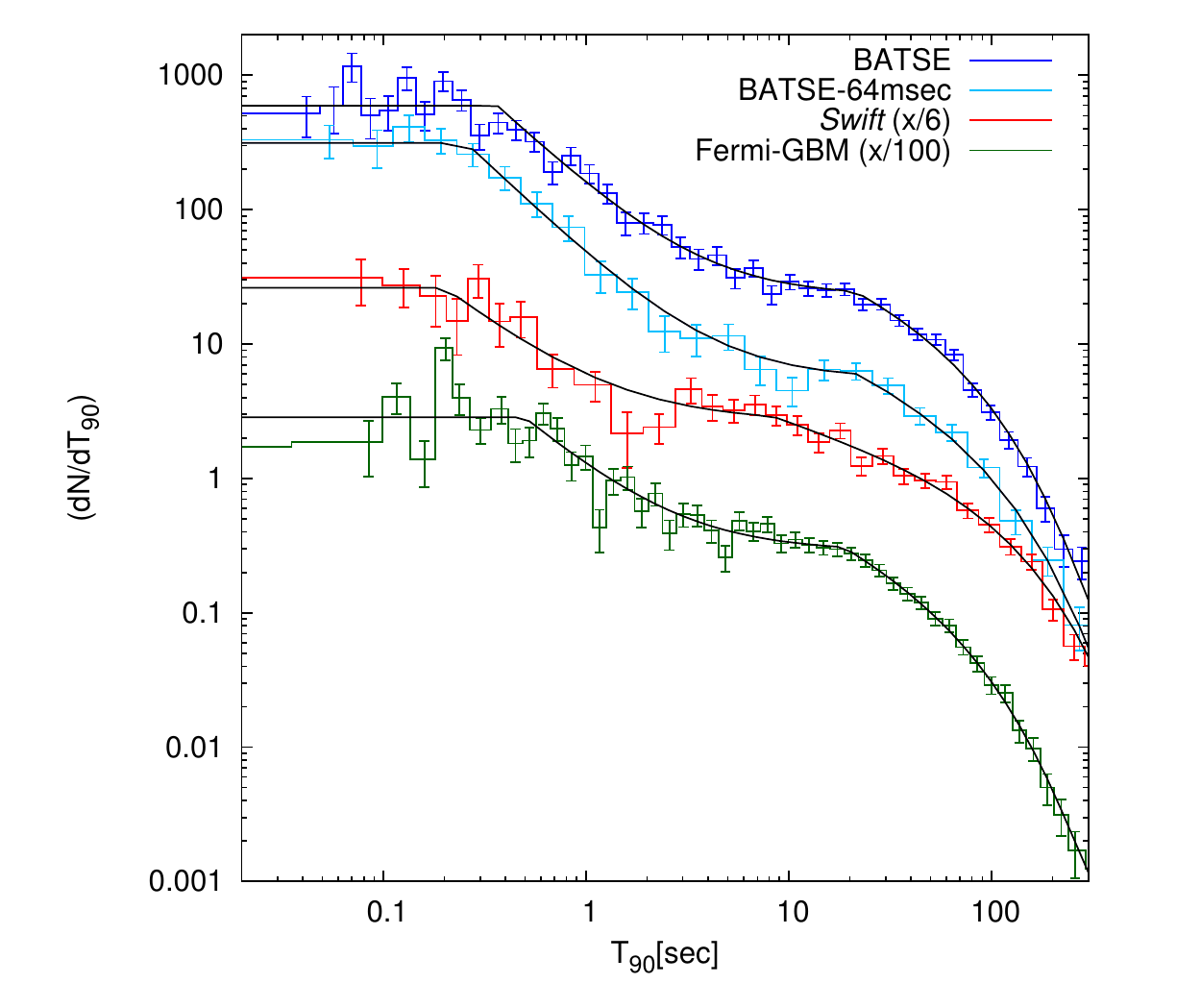}\par
 \includegraphics[width=1.\linewidth]{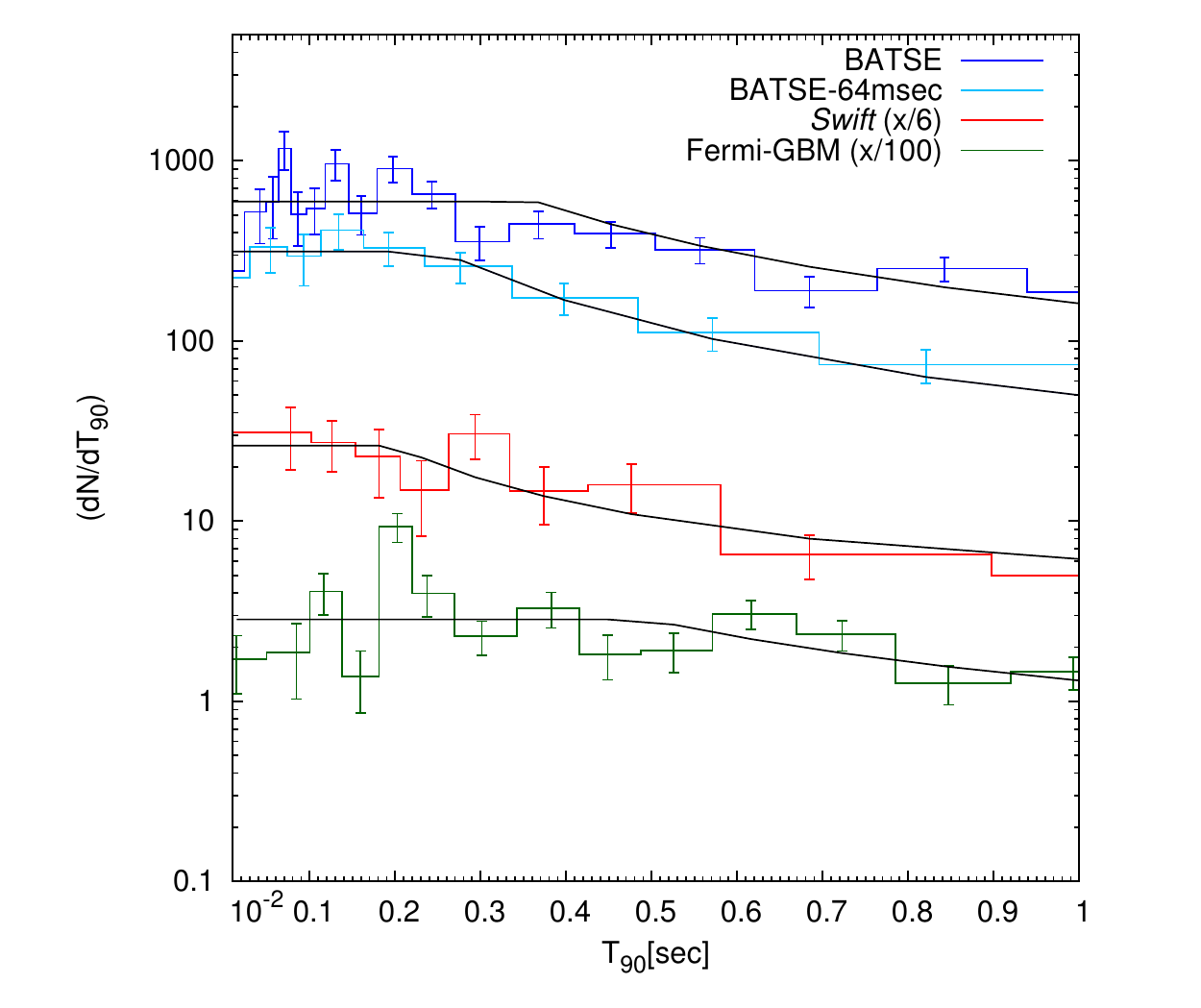}\par
   \end{multicols}
   \caption{Distribution of GRB with respect to $T_{90}$, $dN/dT_{90}$ of BATSE (blue), BATSE$_{64}$ (cyan), {\it Swift}  (red) and Fermi-GBM (green), binned into equally spaced logarithmic bins. Combined best fit for the Collapsar and merger models with black solid lines for all the detectors. The distributions are shown over all range of $T_{90}$ and only for sGRBs with $T_{90}<1$ sec on left and right panel respectively. }
 \label{num_bin}
\end{figure*}

\section{Jet propagation in the ejecta}
\label{theory}
The Collapsar as well the merger models involve jet propagation in the surrounding matter. The relativistic jet forms a double shock structure at its head in which the jet's energy is dissipated. This goes on until the jet's head   breaks out. The jet produces the prompt $\gamma$-emission only once it is far outside the surrounding matter. In both cases the jet is choked and a burst does not arise  if the engine driving it stops operating before the jet's head breaks out. 
An important difference between a Collapsar and a merger is that in the former the surrounding stellar atmosphere is static, while in the latter the ejecta is dynamic and it  expands at mildly relativistic velocity. 
B12 estimate the jet breakout time for a static configuration as: 
 of $t_{b}$  as,  
\begin{equation}
t_{b}=0.4 \, \text{sec}\left( \frac{L_{iso,j}}{10^{51} \text {ergs}/\text {sec}}\right)^{-1/3}\left( \frac{\theta_{j}}{15^{\circ}}\right)^{2/3}\left( \frac{R_{e}}{10^9 \text{cm}}\right)^{2/3}\left( \frac{M_{e}}{10^{-2}M_{\odot} }\right)^{1/3} \  
\label{tbm}
\end{equation}
where   $L_{iso,j}$ is the isotropic equivalent jet luminosity, $\theta_{j}$ the initial jet opening angle,  $M_{e}$ the mass of the surrounding matter and $R_e$ its radius at the time of the jet breakout. Numerical simulations show that this formula is valid  within a factor of order unity. As the ejecta in a merger is expanding one should use a slightly modified formula and now the breakout time is given by the condition: 
\begin{equation}
 \int_0^{T_b} (\beta_h (t) - \beta_{max}) dt = \beta_{max} \Delta t \ , 
 \label{jet_cond}
\end{equation} 
where $\beta_h (t)$ is the jet's head velocity, $\beta_{max}$ is the velocity of the outermost (fastest) shell and $\Delta t$ is the time difference between 
the onset of the jet and the launch of the outflow producing the ejecta. As many of the relevant parameters (e.g. $\Delta t$) are uncertain, we won't attempt to derive an exact formula for the jet breakout time from a merger ejecta and we will use equation \ref{tbm} as a crude estimate. 
Numerical simulations of jet propagation within NS-NS mergers's ejecta  
\citep{Nagakura:2014hza,Murguia-Berthier:2014pta,ore2017} gives values that are slightly smaller but of the same order as equation \ref{tbm}. 

\section{Data}
\label{Data}
We considered the $t_{90}$ duration data from three GRB detectors: BATSE,  {\it Swift}  and {\it Fermi}-GBM.  
These samples have 
$\textbf{N}=2041,1036,1944$ detected bursts  by BATSE {\footnote{https://gammaray.nsstc.nasa.gov/batse/grb/catalog/4b/index.html}}, from 1991 April 21 until 2000 August 17, {\it Swift} {\footnote{https://swift.gsfc.nasa.gov/archive/grb\_table/}}, from 2004 December 17 until 2016 November 5 and Fermi{\footnote{https://heasarc.gsfc.nasa.gov/W3Browse/fermi/fermigbrst.html}}, from 2008 July 17 until 2016 October 22, respectively.

For the BATSE  sample we considered naturally all detected bursts. However,   in  order to check for systematic effects we also considered a 64 msec sub-sample \citep{Schmidt:2001uz}. This sub-sample included bursts that have been triggered in the 64 msec time intervals (i.e. $(C_{max}/ C_{min})_{64{\rm msec}} \ge 1 $, where $(C_{max})_{64{\rm msec}}$ and $(C_{min})_{64{\rm msec}}$ are the peak count rate and the threshold count rate in the 64 msec interval.). $(C_{max}/C_{min})_{64{\rm msec}}$ is  available only for 912 GRBs (from lunch until 29 August 1996) out of the 2041 GRBs in the BATSE catalog. The criterion $(C_{max}/ C_{min})_{64{\rm msec}} \ge 1$ reduces the number of GRBs in this sub-sample to  $\textbf{N}=607$. We denote this sub-sample as BATSE$_{64}$. 

\section{Method}
\label{likeli_method}
Following B12,B13 we have compared  the GRB distribution ($dN_{GRB}/d(T_{90})$) with a theoretical model. Those authors considered only a single ``long"  plateau and fitted the short duration part using a lognormal distribution.  Our theoretical model 
includes both a ``long" Collapsar plateau extending up to $t_{b,C}$
and a ``short" merger plateau extending up to $t_{b,M}$:
\begin{eqnarray}
{p(T_{90)} }= A_{M} \left\{ \begin{array}{ll}
1   \hspace{3. cm}  &T_{90} < t_{b,M}\\                                            
\left( \frac{T_{90}}{t_{b,M}} \right)^{\alpha_{M}}  \hspace{1.7cm} & T_{90} > t_{b,M}\\
\end{array} \right. \nonumber\\
+A_{C}\left\{ \begin{array}{ll}
1   \hspace{3.2 cm}  &T_{90} < t_{b,C}\\                                            
\left( \frac{T_{90}}{t_{b,C}} \right)^{\alpha_C} e^{-\beta_C (T_{90} -t_{b,C})}  \hspace{0.3cm} & T_{90} > t_{b,C} \ .  
\end{array} \right.
\label{distribution}
\end{eqnarray}
Above the jet breakout time  we assume, lacking a better guess, that  the merger distribution follows a power law (with an index $\alpha_M$)  while for the Collapsar distribution we adopt (following B13) a product of a power law (with an index $\alpha_C$ and  an exponential cutoff (characterized by a time scale $1/\beta_C$) \footnote{The  exponential cutoff is phenomenological.  It gives a better fit to the data.}.  $A_{M}$ and $A_{C}$ describe (after normalization) the total number of merger and Collapsar events.

Using a maximum likelihood method,  we fitted this expression to the four data sets. 
The likelihood function  for an individual detector is:
\begin{equation}
 \mathcal{L}=\displaystyle\prod_{i=1}^{\textbf{N}} \left(\frac{p(T_{90,i})}{\int_{T_{90}^{min}}^{T_{90}^{max}}p(T_{90}) dT_{90}}\right).
 \label{ind_lik}
\end{equation}
$T_{90}^{min}$ and $T_{90}^{max}$ has been taken from the GRB samples for the respective detector. 
For each one of the detectors we have maximized  the likelihood function  for the 6 parameters ($A_{M}/A_{C}$, $t_{b,M}$, $\alpha_{M}$, $t_{b,C}$, $\alpha_{C}$, $\beta_{C}$) simultaneously.

We have also carried out a joint analysis for the combined 
likelihood function for the three samples:
\begin{eqnarray}
 \mathcal{L}=&\displaystyle\prod_{i=1}^{\textbf{N}_{\textbf{BATSE}}} \left(\frac{p(T_{90,i})}{\int_{T_{90}^{min}}^{T_{90}^{max}}p(T_{90}) dT_{90}}\right) \times
  \displaystyle\prod_{i=1}^{\textbf{N}_{\textbf{Swift}}} \left(\frac{p(T_{90,i})}{\int_{T_{90}^{min}}^{T_{90}^{max}}p(T_{90}) dT_{90}}\right) \nonumber  \\ & \times  \displaystyle\prod_{i=1}^{\textbf{N}_{\textbf{Fermi}}} \left(\frac{p(T_{90,i})}{\int_{T_{90}^{min}}^{T_{90}^{max}}p(T_{90}) dT_{90}}\right) .
  \label{joint_lik}
\end{eqnarray}
When carrying out the joint analysis we  kept the same $t_{b,M}$ and $\alpha_{M}$, that describe the mergers' distribution,  for all the detectors. However, 
 $T_{90}$ has been observed in different energy bands for the three detectors, [50-300] keV \citep{meegan96}, [15-150] keV \citep{Barthelmy:2005hs} and [50-300] keV \citep{vonKienlin:2014nza} for BATSE, {\it Swift}  and Fermi-GBM respectively. As Collapsar and merger bursts have different spectrum (the latter are harder) this has led to different ratios of ``short" to ``long" bursts in the different detectors. To allow for that we have carried out 
the joint analysis by maximizing the logarithm of the likelihood function for 14 parameters in total, representing 5 different parameters for each detector. Note that in particular we allow the ratio $A_M/A_C$ to vary from one detector to another.

\section{Results }
\label{ana_result}
The best fit parameters and their $1\sigma$ uncertainty are listed in Table~\ref{table_fit}. 
Figure~\ref{num_bin} depicts the GRB duration distributions for equally spaced logarithmic bins of $T_{90}$ for  the different detectors and the best fit
obtained with our model.  The ``short" plateau is clearly seen. It is shown more clearly in the right hand panel of this figure that depicts the region $T_{90} < 1 $ sec.  
The figure contains the histogram of $dN/dT_{90}$  in case of the detectors individually. The standard deviation of the fitting from data can be understood from $\chi^2/d.o.f$, (72.2/42), (35.3/28) and (87.5/46)  for BATSE, {\it Swift}  and Fermi-GBM respectively.

The parameters, we find here,  for the Collapsar model ($t_{b,C}$, $\alpha_{C}$ and $\beta_{C}$) are consistent with the previous analysis of B13 within the statistical uncertainty.  This consistency is reassuring as { we have used unbinned maximum likelihood analysis whereas B13 used a  least squared fit for a binned duration distribution. Again, our analysis contain a factor of two more  bursts compared with   that used previously for Fermi-GBM and {\it Swift}}.  
The best fit values of  the ``long" plateau  are $t_{b,C} = 21.3\pm0.9, \,18.6_{-1.5}^{+0.9} \, \rm{and}\, 8.5_{-1.6}^{+1}$ sec for BATSE, Fermi-GBM and {\it Swift} respectively. The discrepancy in the values for different detectors is natural  as  $T_{90}$ is observed  for different energy bands and with different time resolutions and triggering criteria. Table~\ref{table_fit} also lists the best fit parameters and their uncertainty for the case of the joint analysis. The ``long" plateau (corresponding to the Collapsar's jet breakout times), $t_{b,C}$, for the  individual detectors  remain the same for this  analysis  as expected.

\begin{figure*}
 \begin{multicols}{2}
  \includegraphics[width=1.0\linewidth]{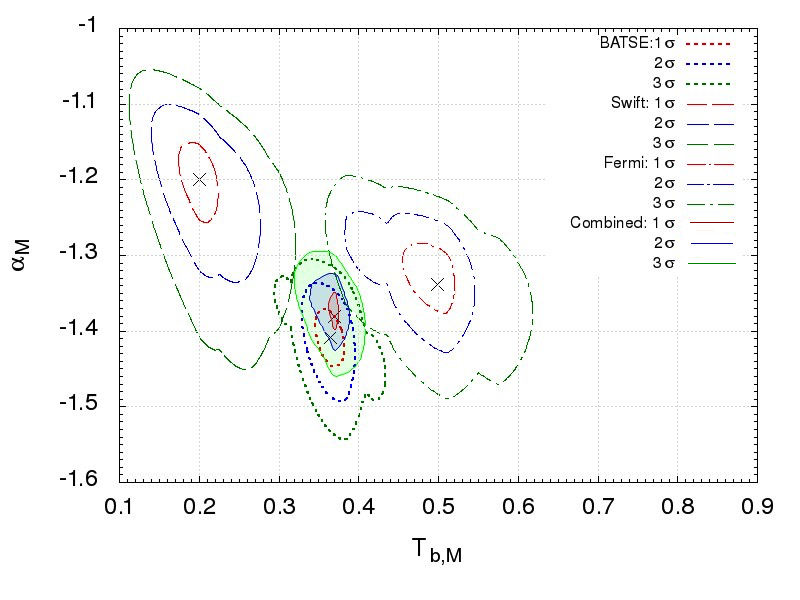}\par
  \includegraphics[width=1.0\linewidth]{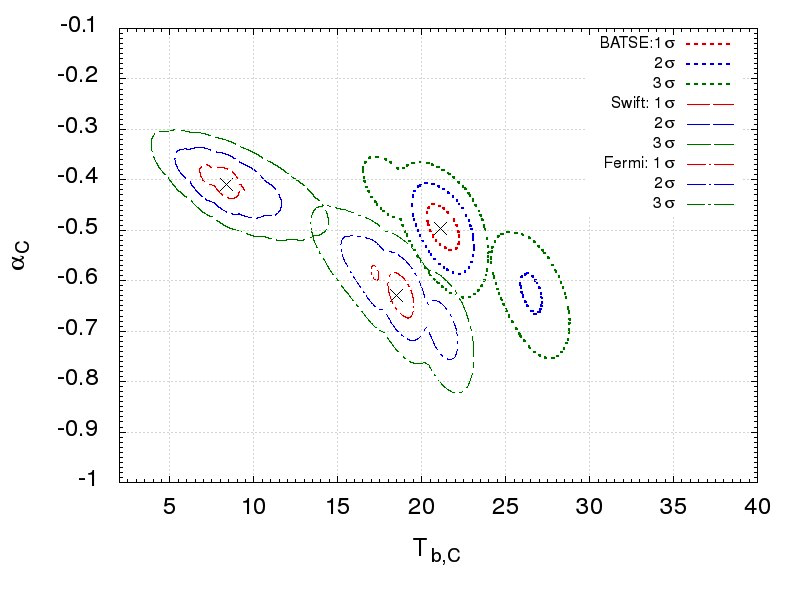}
\end{multicols}
    \caption{Left: Contour plots for the merger parameters, $t_{b,M}$ and $\alpha_{M}$ for BATSE (dotted lines), Fermi (dot-dashed lines) and {\it Swift}  (dashed lines) and combined analysis of BATSE, Swift and Fermi (solid lines). The contours are for likelihood ratios corresponding to $1\sigma$ (red color), $2\sigma$ (blue), $3\sigma$ (green). The best fit values are marked with black cross. Right: Same as left for the Collapsar parameters, $t_{b,C}$ and $\alpha_{C}$. }
\label{figtest1}
\end{figure*}

Turning now to  the  short duration part of the $T_{90}$ distribution   we find 
the best fit of  $t_{b,M}= 0.37\pm0.02,\, \, 0.5\pm0.02 \, \rm{and} \, 0.2\pm0.03,$ sec for BATSE, Fermi-GBM and {\it Swift}  respectively. For the joint analysis  we find $t_{b,M}=0.37\pm0.006$ sec.  
The likelihood contour plots of $t_{b,M}$ vs $\alpha_{M}$ and $t_{b,C}$ vs $\alpha_{C}$ in case of individual detectors are shown in the left and right panel of Figure~\ref{figtest1} respectively. The contours represent likelihood ratio $e^{-0.5}$, $e^{-2}$, $e^{-4.5}$, corresponding to $1\sigma$, $2\sigma$, $3\sigma$ uncertainty range respectively. 
 
The best fit values for the BATSE$_{64}$ sample  give further  support to our findings. 
For this sample we find   $T_{b,M} = 0.25\pm0.02$ sec. The other parameters related to the merger model from this analysis are, $A_M/A_C= 56\pm5$, $\alpha_{M}=-1.44_{-0.23}^{+0.03}$. Naturally  ratio of mergers to Collapsars is larger here and this leads to a somewhat less pronounced  ``long" plateau. Still the end of the ``long" plateau is at $t_{b,C} =21.4_{-1}^{+7.4}$ sec - just like in the full BATSE sample. Other best fit values $\alpha_{C}=-0.5_{-0.2}^{+0.04}$ and 
$\beta_{C} =0.012\pm0.0002$ are also consistent.

\begin{table*}

\begin{multicols}{2}
\centering
\fontsize{10}{9}
\centering
\begin{tabular}{c||c|c|c|c|c|c}
\hline
 &$A_{M}/A_C$  &  $t_{b,M}$ & $\alpha_{M}$ & $t_{b,C}$ & $\alpha_{C}$ & $\beta_C$\\
 \hline
 \hline
BATSE &  $25\pm 1.5$  & $0.37\pm0.02$  &  $-1.4\pm0.04$    &  $21.3\pm0.9$  & $-0.5\pm 0.056$  & $0.0156 \pm 0.0004$  \\
BATSE$_{64}$ &  $56\pm 5$  & $0.25\pm0.02$  &  $-1.44^{+0.04}_{-0.023}$    &  $21.4^{+7.4}_{-1} $  & $-0.5^{+0.04}_{-0.2}$  & $0.012 \pm {0.0002}$  \\
{\it Swift} &  $8.9\pm 1.2$  & $0.2\pm0.03$   & $-1.2\pm0.05$    &  $8.5^{+1}_{-1.6}$  &  $-0.4\pm0.035$   &  $0.009 \pm 0.0004$ \\
Fermi &  $9\pm 0.75$  & $0.5\pm{0.02}$   & $-1.33\pm0.045$    &  $18.6^{+0.9}_{-1.5}$  &  $-0.62\pm{0.45}$   & $0.015 \pm 0.001$ \\
\hline
\multicolumn{7}{ c }{Joint Analysis}\\
\hline
BATSE &  $25.4^{+0.6}_{-2.6}$  &  {${0.37\pm0.006}$}  &  ${-1.38\pm0.03}$ &  $21.4^{+6.6}_{-0.9}$  & $-0.5^{+0.2}_{-0.01}$  & $0.016 \pm 0.0007$  \\
{\it Swift} &  $6.8\pm 1.2$  &  &   &  $8.6\pm1$  &  $-0.4\pm{+0.01}$   &  $0.009 \pm 0.0005$ \\
FERMI &  $10.7\pm 0.8$  &  &    &  $18.6^{+0.8}_{-2.2}$  &  $-0.75\pm0.05$   & $0.0133 \pm 0.0007$ \\
\hline
\end{tabular}
\end{multicols}
 \caption{The best fit values of the parameters and their uncertainty within $2\sigma$ for individual detectors (upper panel) and combined analysis (lower panel).}
 \label{table_fit}
\end{table*}

\begin{table}
\fontsize{6}{7}
\centering
\begin{tabular}{c||c|c|c|c}
 \hline
 & - $log \mathcal{L}^{Max}_{plateau} $ & - $log\mathcal{L}^{Max}_{log-normal} $ & $\sqrt{\mathcal{TS}}$ & P-value\\
  \hline
 \hline
 BATSE & 8817.31 & 8820.34 & 2.46 & 0.014 \\
 Swift & 5326.40 & 5335.32 & 4.22 & $2.4\times10^{-5}$\\
 Fermi & 8746.28 & 8747.63 & 1.64 & 0.1\\
 \hline
\end{tabular}
 \caption{A likelihood ratio test showing that  a plateau is preferred over  a  log-normal distribution  when fitting the sGRBs duration distribution. }
 \label{table_com}
\end{table}

B13 fitted the  short duration population with a log-normal distribution in which the first part of equation~\ref{distribution}
is replace by log-normal expression, $\frac{1}{T_{90}\sigma\sqrt{2\pi}}\rm{exp}\left({-\frac{(lnT_{90}-\mu)^2}{2\sigma^2}}\right)$.  We have compared this log-normal model to the  plateau model (equation~\ref{distribution}) using a similar likelihood analysis for the  individual detectors. 
The best fit values for the long normal model agree with those derived by B13, even though for {\it Swift} and Fermi-GBM the data set is twice as large. 
The maximum likelihood function for the log-normal model, $\mathcal{L}^{Max}_{log-normal}$ and the one for the plateau model, $\mathcal{L}^{Max}_{plateau}$,  are  listed in Table~\ref{table_com}. The corresponding likelihood ratio (Test Statistic), $\mathcal{TS} = - 2 \, log\left({\mathcal{L}^{Max}_{log-normal}}/{\mathcal{L}^{Max}_{plateau}}\right)$, is listed as well. The number of parameters in both distributions are the same so the d.o.f for $\mathcal{TS}$ is 0. The p-value (considering Gaussian standard deviations $\sigma=\sqrt{\mathcal{TS}}$), represents the probability that the GRB distribution ($dN/dT_{90}$) follows a plateau model over a log-normal model.

Following B13 we define as a  useful threshold duration that separates Collapsars from non-Collapsars  at the duration for which $f_{C}(T_{th,90})=0.5$, i.e. the fraction,  $f_{C}(T_{90})$, of Collapsars out of the total number of bursts is a half (see equation~\ref{distribution}). 
Below this value a burst is more likely to be a merger. Above this value a burst is more likely to be a Collapsar\footnote{B13 suggest a more refined probability distribution that depends on the hardness of the bursts. }. 
We find the transitions time as: $T_{th,90}=3.6^{+0.3}_{-0.1}, \,  1.2\pm0.2$, and  $2.6\pm0.2$ sec for BATSE, {\it Swift}  and Fermi-GBM  respectively. The first two values are within the $1\sigma$ range of the corresponding values found by B13, $3.1\pm 0.5$ and  $0.8\pm0.3$ sec. The last one (for Fermi-GBM) is farther away: $1.7\pm0.4$   vs. $2.6\pm0.2$ sec but the difference is reasonable given the fact that we are fitting a different functional form to the duration distribution. Note that the overall trend of smaller values of $T_{th,90}$ 
found for the log-normal distribution is consistent with the fact that this distribution falls more steeply than the power-law distribution that we are using above the plateau.

\section{Discussion}
\label{discussion_con}

We have examined the duration distribution of prompt GRB emission for BATSE, Fermi-GBM and {\it Swift} using a maximum likelihood analysis. The distribution is well fitted by a two plateaus one with a short duration and the other with a longer duration. These plateaus are consistent with those expected from models in which a jet is propagating in a stellar atmosphere (for Collapsars; B12,B13) and with a jet propagates in an ejecta (for mergers).   
In the earlier least square fit analysis done by B13, the best fit for the duration of the ``long" plateau , $t_{b,C}=19.4_{-4.2}^{2.5}$ sec for BATSE. For this analysis the data is the same as in our analysis and indeed we find a similar value: $21.3 \pm 0.9$.
This value remain unchanged within 1$\sigma$ error with our  maximum likelihood analysis. 
The only difference concerning the BATSE data is the different fit at short durations. 
For {\it Swift}  and Fermi-GBM the data in B13 was taken till 2012, however in our analysis we have taken all bursts until 2016,  increasing the number of data points by a factor of two. 
Still we obtained similar results  to B13. %The other parameters related to Collapsar model, $A_{M}/A_{C}$ and $\beta_C$ remain similar to the analysis done by B13.
The  consistency of the  plateaus in the long bursts distributions that are verified with twice the  data and with a different statistical analysis gives a concrete and statistically significant support to LGRB progenitors as Collapsars. The jet breakout time of around few dozen of seconds, predicted for standard Collapsar model for LGRBs
(B12)  is consistent with the best fit value obtained for the three detectors. 

More importantly we have found a 
 plateau also in the short duration part of the data with $t_{b,M} \sim 0.2 -0.5$ sec for the three detectors.  
 The difference in $t_{b,M}$ for different detectors can be easily understood in view of their different detection windows and triggering procedure  \citep[see e.g.][for differences between BATSE and GBM and {\it Swift } and BATSE respectively] {2011MNRAS.415.3153N,2009A&A...504...67H}.
% , the short bursts detected by Fermi-GBM has $E_{peak}^{obs}$ (peak energy) larger compared to BATSE events, leading to the fact that GBM might miss sGRBs at low fluences. The other interpretation by \citep{} from the detection ability {\it Swift}  sGRBs can be lower in number compared to BATSE.
 For all detectors the ``short" plateau model gives a better fit to the data than 
 the log-normal model used in the earlier analysis. The duration of the plateau is consistent with the predicted jet breakout time provided that the jet  propagates in a mass ejecta of order $ 0.01M_{\odot}$ that is expanding with a velocity of order 0.1c, as expected from modeling of outflows in neutron star mergers. Interestingly the existence of the plateau also suggest that there are merger events in which the central engine didn't operate long enough for the jets to breakout and produce a short GRB.

This analysis indicates that short GRBs jets penetrate, before breaking out, a surrounding mass of order of a percent of a solar mass strongly support the neutron star mergers model for short GRBs \citep{Eichler:1989ve}. . Numerical simulations of the merger process suggest that mass of this order of magnitude is ejected. This is consistent with indications of macronovae accompanying short GRBs \citep{Berger2013,tanvir2013,Yang:2015pha,Jin:2015dxh}. Such macronovae arise from this ejecta and they provide a comparable estimates of the ejecta masses. A further confirmation of this model would arise from a direct discovery of the emission of the cocoon that is produced when the jet propagates through the ejecta \citep{ore2017}.

\section*{Acknowledgements}
We  thank Kenta Hotokezaka and Ehud Nakar for helpful discussions. This research was supported by the I-Core center of excellence of the CHE-ISF, by  an advanced ERC grant TREX and by a grant from the Templeton foundation.

%%%%%%%%%%%%%%%%%%%%%%%%%%%%%%%%%%%%%%%%%%%%%%%%%%

%%%%%%%%%%%%%%%%%%%%%%%%%%%%%%%%%%%%%%%%%%%%%%%%%%

% Don't change these lines
\label{lastpage}
\end{document}